\documentclass[aps,prd,preprint,superscriptaddress]{revtex4-1}
\usepackage{amsmath,amssymb}
\usepackage{bm}
\usepackage[dvips]{graphicx}
\usepackage{here}
\usepackage{multirow}
\begin{document}
\title{Relaxation of hierarchy in higher-dimensional Starobinsky model}
\author{Yu Asai}
\email[E-mail:]{u-asai.physics@ruri.waseda.jp}
\affiliation{Department of Physics, Waseda University, Tokyo 169-8555, Japan}
\begin{abstract}
Starobinsky model, which has a Ricci scalar squared term $R^{2}$ in its action, is one of the most promising inflation models from the viewpoint of Cosmic Microwave Background observations. However, it is well known that observational predictions of this model are quite sensitive to the existence of $R^{m}$ $(2<m)$ terms, whose absence is just assumed. In this paper, we clarify that the observational predictions of $D$-dimensional ($4<D$) extended Starobinsky model are less sensitive to such terms than those of the original 4-dimensional model.This result make it easier to construct Starobinsky-like models in higher dimensions.
\end{abstract}
\maketitle
\section{Introduction}
Cosmological inflation\cite{Starobinsky'80,Linde'83,Albrecht and Steinhardt'82} has become a more and more attractive paradigm of cosmology. It can solve initial problems in standard Big Bang cosmology, such as the horizon problem and the flatness problem. In addition, primordial fluctuations, which are seeds of both the large-scale structure of the universe and anisotropy of Cosmic Microwave Background (CMB), can also arise during inflation. Furthermore, along with the recent great progress of cosmological observations, it has become possible to judge which inflation model is favored by the CMB observations\cite{Planck2015,Planck2018}. 
\par
Starobinsky model\cite{Starobinsky'80} is one of the most promising inflation because its predictions fall into the center of CMB observational constraints\cite{Planck2018}. A distinctive feature of Starobinsky model which we have to remark is that it contains Ricci curvature squared term in the action;
\begin{align}
\frac{M_{P}^{2}}{2}\int d^{4}x\sqrt{-g}\Bigr(R+\frac{1}{2M^{2}}R^{2}\Bigl).\label{Starobinsky action intro}
\end{align}
where $M_{P}$ is 4-dimensional Planck mass and $M$ is a parameter which has mass dimension 1. Although it is important to investigate the origin of the higher curvature term for high energy physics, however, the origin is still unknown. One interesting direction to unravel it is to regard such higher curvature term as a part of an effective action of high energy physics. For instance, a gravity part of string/M theory effective action is expected to appear as follows (see \cite{Bento and Bertolami'95} and references with in);
\begin{align}
\frac{M_{(D)}^{D-2}}{2}\int d^{D}x\sqrt{-g}\biggr(R+\sum_{m=1}^{\infty}\mathcal{L}^{(m)}(R_{\mu\nu\rho\sigma},g_{\mu\nu})\biggl),\label{a series of curvature}
\end{align}
where $M_{(D)}$ is $D$-dimensional Planck mass and $\mathcal{L}^{(m)}$ denotes a part of Lagrangian which contains $m$-th order curvature. Note that $\mathcal{L}^{(m)}$ is not Lovelock Lagrangian\cite{Lovelock'71} in general, hence whether or not ghost modes\cite{Ostrogradsky1850} arise from higher derivative terms in the model greatly depends on the detailed form of $\mathcal{L}^{(m)}$. In the case when models contain the ghost modes, it is said that such models are unstable and physically unacceptable.
\par
Therefore, we will consider the following model instead of Eq.(\ref{a series of curvature}) in this paper since it is well known that the action which contains only Ricci scalar curvature has no ghost mode;
\begin{align}
\frac{M_{(D)}^{D-2}}{2}\int d^{D}x\sqrt{-g}\biggr(R+\sum_{m=1}^{\infty}\frac{\lambda_{m}}{mM^{2m-2}}R^{m}\biggl),\label{a series of curvature toy model}
\end{align}
where $\lambda_{m}$ is a dimensionless parameter and $M$ is a mass scale above which higher curvature effects become important. The $D$-dimensional action (\ref{a series of curvature toy model}) can be considered as a higher-dimensional generalization of the Starobinsky action Eq.(\ref{Starobinsky action intro}), including  higher curvature terms, $\lambda_{m} R^{m}/mM^{2m-2}(2<m)$.
\par
In Ref.\cite{Qing-Guo Huang'14}, the author added a $\lambda R^{m}/mM^{2m-2}(2<m)$ term to 4-dimensional Starobinsky action (\ref{Starobinsky action intro}) and estimated its effects on observational predictions. Then they obtained severe constraints on $\lambda$ from CMB observations, which require a large hierarchy between a $R^2$ term and other higher curvature terms. This hierarchy makes it difficult to consider the high energy origin of the Starobinsky action. On the other hand, in Refs.\cite{Ketov and Nakata'17a,Otero et al'17,Ketov and Nakata'17b}, the authors extended the Starobinsky model to a $D$-dimensional one. In those papers, they considered $R+R^{n}/nM^{2n-2}$ models in $D$-dimensions and concluded $D=2n$ model can cause the inflation whose prediction fits current CMB observations.  
\par
In the present paper, we combine the ideas of the previous papers, i.e., we consider $R+R^{n}/nM^{2n-2}+\lambda R^{m}/mM^{2m-2}(D=2n, m\neq n)$ model in $D$-dimensions and estimate the effects of the $\lambda R^{m}/mM^{2m-2}$ term on observational predictions analytically and numerically. As a result, we find out that constraints on $\lambda$, i.e., the hierarchy among the higher curvature terms is relaxed compared with the original $4$-dimensional model. 
\par
This paper is organized as follows. In Section 2, a review of Starobinsky model and its higher-dimensional extension is given. Although most parts of this section are reviews of previous papers, a discussion on $D\neq 2n$ model has a novelty. Section 3 is the main body of this paper. In this section we clarify the relaxation of hierarchy in higher- dimensional extend Starobinsky model analytically and numerically. Section 4 is devoted to the summary and discussion. Some detailed calculations are shown in Appendix..
\par
Note that in this paper we use natural unit $\hbar=c=1$. The sign convention of the mtirc is chosen as $(-,+,+,+,\cdots)$ and the Ricci tensor is defined as $R_{\mu\nu}={R^{\rho}}_{\mu\rho\nu}$.
\section{Starobinsky model and its higher-dimensional extension}\label{staro D extension}
In this section, we briefly review Starobinsky model and its higher-dimensional extensions. Possibility of this extensions is originally pointed out in the early days of F(R) gravity \cite{Maeda'89,Barrow and Cotsakis'88} and recently a more detailed discussion, including compactification of extra dimensions, is proposed \cite{Ketov and Nakata'17a,Otero et al'17,Ketov and Nakata'17b}.
\subsection{4-dimensional Starobinsky model}
As we mentioned in Introduction, Starobinsky model has a curvature squared term in its action, instead of a scalar degree of freedom, in addition to Einstein-Hilbert action:
\begin{align}
\frac{M_{P}^{2}}{2}\int d^{4}x\sqrt{-g}\Bigr(R+\frac{1}{2M^{2}}R^{2}\Bigl)\label{Starobinsky action},
\end{align}
 From the observation of CMB power spectrum, one must tune the parameter $M\sim \mathcal{O}(10^{-5})\times M_{P}$. Of cause, this is also a kind of hierarchy but it is not an aim of this paper. We will come back, however, to this point in the summary and discussion section. 
\par
Starobinsky model is included in $F(R)$ gravity theories (see Ref.\cite{Tsujikawa review} for a review). One can obtain Eq.(\ref{Starobinsky action}) by appropriately choosing a function form of $F(R)$. Therefore, as all $F(R)$ gravity theories have, Starobinsky model has a higher derivative degree of freedom. One can recast Eq.(\ref{Starobinsky action}) in a form where the degree of freedom appears explicitly, by introducing a Lagrange multiplier field, applying Weyl transformation and redefining fields:
\begin{align}
\int d^{4}x\sqrt{-g}\Bigr(\frac{M_{P}}{2}R-\frac{1}{2}g^{\mu\nu}\partial_{\mu}\phi \partial_{\nu}\phi-V(\phi)\Bigl), \hspace{1em}V(\phi)\equiv-\frac{1}{4}M^{2}M^{2}_{P}\bigr(1-e^{-\sqrt{\frac{2}{3}}\phi/M_{P}}\bigl)^{2},\label{4 dim Starobinsky}
\end{align}
where $\mu$, $\nu$ runs 4-dimensional spacetime. This action, which is often called dual form of Starobinsky action, has Einstein-Hilbert term and a scalar field which has canonical kinetic term and extremely flat scalar potential. This scalar field $\phi$ represents the higher degree of freedom and is called ``scalaron''. This shape of scalar potential predicts the following spectral index $n_{s}$ and tensor-to-scalar ratio $r$, if one identifies this scalaron as inflaton:
\begin{align}
n_{s}\simeq1-\frac{2}{N_{e}}+\cdots, \hspace{3em} r\simeq\frac{12}{N_{e}^{2}}+\cdots,
\end{align}
where $\cdots$ denotes higher order contributions of slow-roll parameters, which we neglect here, and $N_{e}$ is e-folding number between horizon crossing and inflation end ($N_{e}=50\sim60$). This prediction nicely fits Planck2018 result\cite{Planck2018}. 
\subsection{$D$-dimensional Starobinsky model}
Recently higher-dimensional extensions of Starobinsky model have been discussed by some researchers, motivated by higher dimensional models of high energy physics. In this subsection, we review $D$-dimensional extensions of Starobinsky model. First, let us consider models whose actions are
\begin{align}
\frac{M_{(D)}^{D-2}}{2}\int d^{D} x\sqrt{-g}\Bigr(R+\frac{1}{nM^{2n-2}}R^{n}\Bigl),\label{d dim Starobinsky F(R) frame}
\end{align}
where $M_{(D)}$ is $D$-dimensional Planck scale and $n$ is an arbitrary integer at this stage. As with 4-dimensional Starobinsky model in the previous subsection, one can recast this actions into dual forms (see the Appendix for the details),
\begin{align}
\frac{M_{(D)}^{D-2}}{2}\int d^{D}x\sqrt{-g}\biggl(R-\frac{D-1}{D-2}g^{AB}\partial_{A}\bar{\phi}\partial_{B}\bar{\phi}-\frac{n-1}{n}M^{2}e^{-\frac{D}{D-2}\bar{\phi}}\bigr(e^{\bar{\phi}}-1\bigl)^{\frac{n}{n-1}}\biggr),\label{D dim Starobinsky in D dim dual form}
\end{align}
where $A$, $B$ runs $D$-dimensional spacetime and $\bar{\phi}$ represents a higher derivative degree of freedom. For the scalar potential to be real, $\bar{\phi}$ must take positive value if $n$ is odd. In previous researches\cite{Otero et al'17,Ketov and Nakata'17b}, the authors discussed compactification of $D$-dimensional spacetime into 4-dimensional spacetime by introducing a form field flux and obtained a 4-dimensional action. In this paper, we just assume the following 4-dimensional actions for simplicity:
\begin{align}
\int d^{4}x\sqrt{-g}\Bigr(\frac{M_{P}^{2}}{2}R-\frac{1}{2}g^{\mu\nu}\partial_{\mu}\phi \partial_{\nu}\phi-V(\phi)\Bigl),\hspace{1em}V(\phi)\equiv\frac{n-1}{2n}M_{P}^2 M^{2}e^{-\frac{D}{D-2}\alpha \phi}\bigr(e^{\alpha \phi}-1\bigl)^{\frac{n}{n-1}},\label{D dim Starobisnky}
\end{align} 
where we neglect dilaton and Kaluza-Klein vector and assume that all fields depend only on 4-dimensional coordinate. Here $\bar{\phi}\equiv\sqrt{\frac{D-2}{D-1}}\frac{\phi}{M_{P}}\equiv\alpha \phi$, $M_{P}^2=M_{(D)}^{D-2}\mathcal{V}_{\rm extra}$ and  $\mathcal{V}_{\rm extra}$ is volume of extra dimensions.
\par
\begin{figure}[htbp]
\centering
\includegraphics[width=10cm]{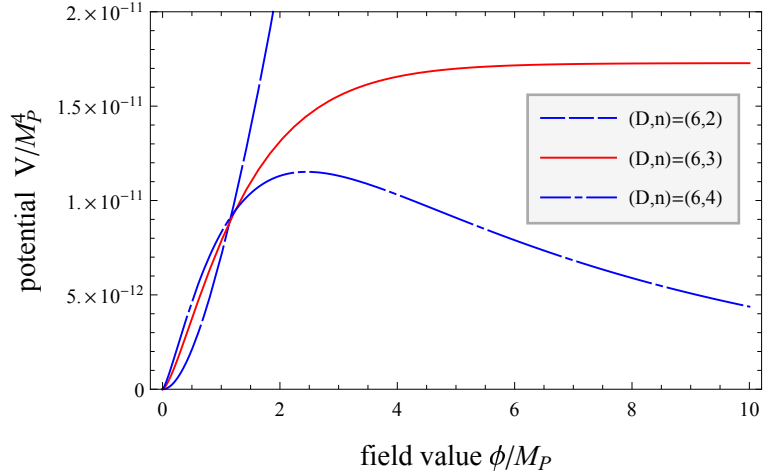}
\caption{This figure shows the shapes of potentials of Eq.(\ref{D dim Starobisnky}) with $(D,n)=(6,2),(6,3),(6,4)$. Here we set $M=7.2\times10^{-6}M_{P}$.}
\label{potential shape}
\end{figure}
\begin{figure}[htbp]
\centering
\includegraphics[width=10cm]{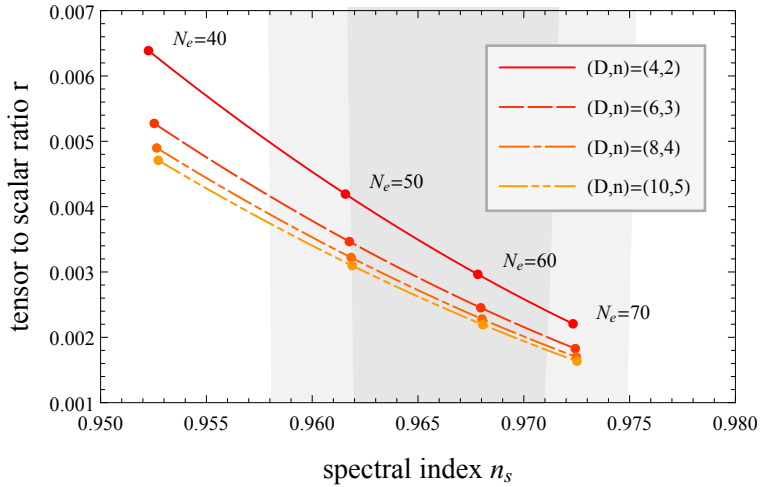}
\caption{This figure shows $n_{s}$-$r$ plot in $D=4,6,8,10$-dimensional Starobinsky models ($D=2n$). The gray region denotes constraints from PlanckTT+lowP+BKP+lensing+ext experiments, which is traced from Fig.18 in Planck2018 results\cite{Planck2018}.}
\label{D=2n ns-r plot}
\end{figure}

The scalar potentials of Eq.(\ref{D dim Starobisnky}) becomes $V(\phi)\propto \exp \alpha (\frac{n}{n-1}-\frac{D}{D-2})\phi$ at large field region ($1\ll\bar{\phi}$). One can thus obtain the following extremely flat scalar potential if $D=2n$:
\begin{align}
V(\phi)|_{D=2n}=\frac{n-1}{2n}M_{P}^2 M^{2}\Bigr(1-e^{-\alpha \phi}\Bigl)^{\frac{n}{n-1}}.\label{D dim Starobinsky potential}
\end{align} 
In $n=2$ case, this scalar potential reproduces Starobinsky potential (\ref{4 dim Starobinsky}). Therefore we call a model with $D=2n$ as $D$-dimensional Starobinsky model. The fact that Eq.(\ref{D dim Starobisnky}) has a flat scalar potential in $D=2n$ can be confirmed explicitly by drawing  the scalar potential (Fig.\ref{potential shape}).
The potential predicts the following spectral index  $n_{s}$ and tensor-to-scalar ratio $r$:
\begin{align}
n_{s}\simeq1-\frac{2}{N_{e}}+\cdots, \hspace{3em} r\simeq\frac{4(2n-1)}{n-1}\frac{1}{N_{e}^{2}}+\cdots,
\end{align}
where $\cdots$ denotes higher order contributions of slow-roll parameters, which we neglect here. As with 4-dimensional Starobinsky model, the predictions nicely fit Planck2018 results (Fig.\ref{D=2n ns-r plot}). A difference of dimensions  appears only in tensor-to-scalar ratio $r$ at leading order of slow-roll parameters. This difference may be detected in future observations.
\par
\begin{figure}[htbp]
\centering
\includegraphics[width=8cm]{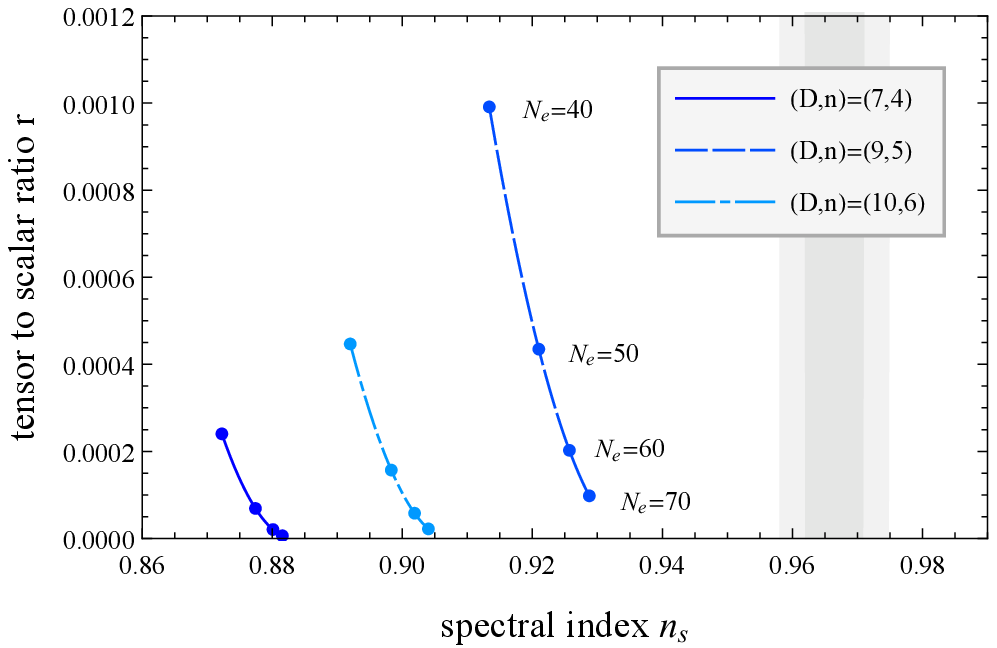}
\includegraphics[width=8cm]{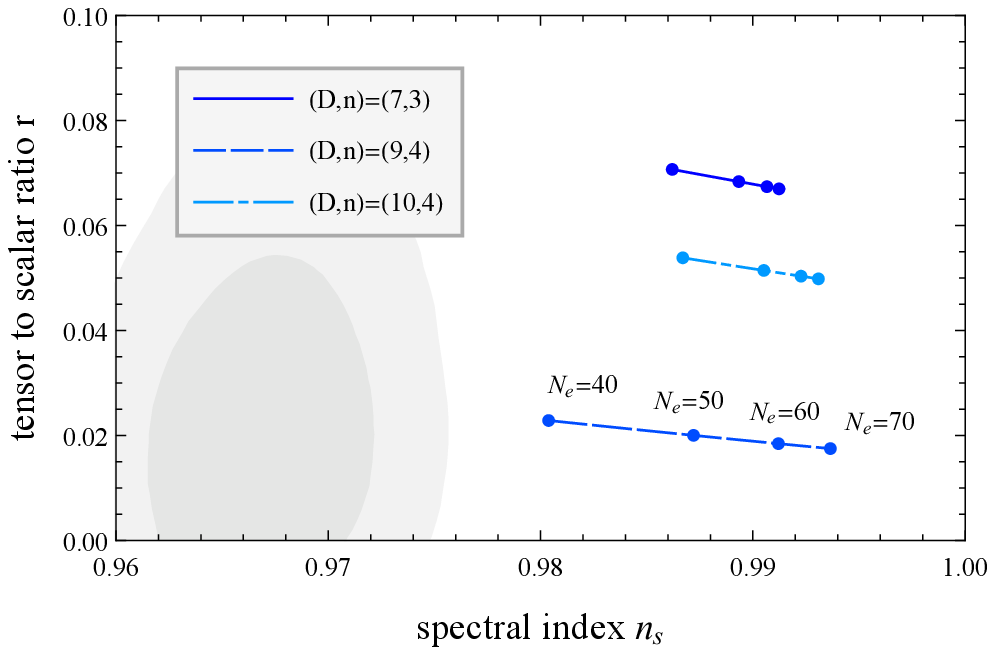}
\caption{These figures show $n_{s}$-$r$ plot in $D<2n$ models (left panel) and $D>2n$ models (right panel) varying $D$ and $n$. The gray region denotes constraints from PlanckTT+lowP+BKP+lensing+ext experiments, which is traced from Fig.18 in Planck2018 results\cite{Planck2018}.}
\label{Dnot=2n ns-r plot}
\end{figure}

One might think inflation successfully works even when $D\neq2n$. However we found that all of $D\neq 2n$ inflation models in $4\le D \le 10$ are rejected by constraints from Planck2018 results\cite{Planck2018} (Fig.\ref{Dnot=2n ns-r plot}). This is generalization of the fact that 4-dimensional $R+R^n/M^{2n-2} (n>2)$ inflation models does not fit the observations\cite{Kaneda Ketov and Watanabe'10,Motohashi'14}.
\section{Relaxation of hierarchy in higher-dimensional Starobinsky model}\label{hierarchy}
$D$-dimensional extension of Starobinsky model can make successful predictions of observation. However, as mentioned in the introduction, there is no reason not to add $R^{m} (m\neq n)$ term to $D$-dimensional Starobinsky model. In this section, we will consider the following models:
\begin{align}
\frac{M_{(D)}^{D-2}}{2}\int d^{D}x\sqrt{-g}\biggr(R+\frac{1}{nM^{2n-2}}R^{n}+\frac{\lambda}{mM^{2m-2}}R^{m}\biggl),\label{D dim Starobinsky + correction}
\end{align}
where $D=2n$, $m\neq n$ and $\lambda$ is a dimensionless parameter. In the following section, we will discuss $\lambda$ dependence on spectral index $n_{s}$ and tensor-to-scalar ratio $r$ when we vary dimensions of spacetime $D$ and power of Ricci scalar in additional term $m$. Also we will estimate allowed range of $\lambda$ in various $D$ and $m$ from Planck2018 results.
\subsection{Analytical approach}
\begin{figure}[htbp]
\centering
\includegraphics[width=10cm]{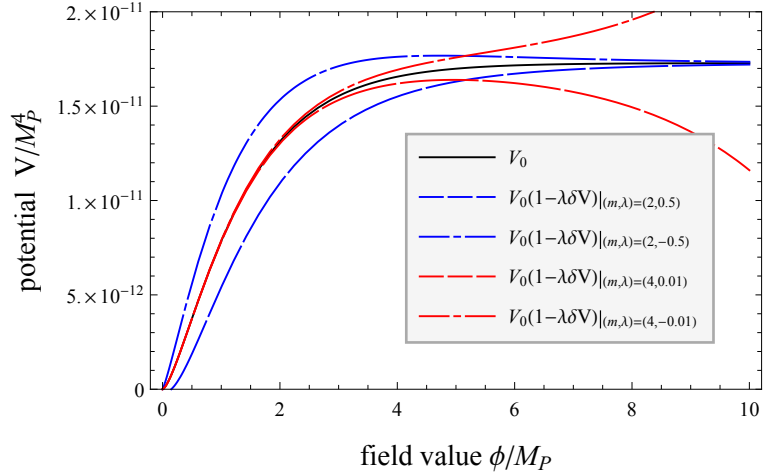}
\caption{This figure shows the shape of potential Eq.(\ref{corrected D dim Starobinsky potential}) in $(D,n)=(6,3)$ case, neglecting higher order terms $\mathcal{O}((\lambda\delta V)^2)$. Here we set $M=7.2\times10^{-6}M_{P}$}
\label{modified potential shape}
\end{figure}

In this subsection, we will discuss $\lambda$ dependence on $n_{s}$ and $r$ in analytical approach. We will extend 4-dimensional method used in previous research\cite{Qing-Guo Huang'14}.  
First, we rewrite Eq.(\ref{D dim Starobinsky + correction}) into dual form (see the appendix for the details),
\begin{align}
&\frac{M_{(D)}^{D-2}}{2}\int d^{D}x\sqrt{-g}\Biggl[R-\frac{D-1}{D-2}g^{AB}\partial_{A}\bar{\phi}\partial_{B}\bar{\phi}\nonumber\\
&\hspace{10em}-M^{2}e^{-\frac{n}{n-1}\bar{\phi}}\Bigl(\frac{n-1}{n}\Bigl(\frac{\chi}{M^{2}}\Bigr)^{n}+\lambda\frac{m-1}{m}\Bigl(\frac{\chi}{M^{2}}\Bigr)^{m}\Bigr)\Biggr].\label{dual form of D-dimensional Starobinsky model with correction}
\end{align}
Here and $\chi$ is a solution of the equation below:
\begin{align}
\biggr(\frac{\chi}{M^{2}}\biggl)^{n-1}+\lambda\biggr(\frac{\chi}{M^{2}}\biggl)^{m-1}=e^{\alpha\phi}-1.\label{chi equation to solve}
\end{align}
As with previous section, we assume the following 4-dimensional actions:
\begin{align}
&\int d^{4}\sqrt{-g}\Bigr(\frac{M_{P}^{2}}{2}R-\frac{1}{2}g^{\mu\nu}\partial_{\mu}\phi \partial_{\nu}\phi-V(\phi)\Bigl),\nonumber\\
V(\phi)\equiv &M_{P}^{2}M^{2}e^{-\frac{n}{n-1}\alpha\phi}\biggl(\frac{n-1}{2n}\biggr(\frac{\chi}{M^{2}}\biggl)^{n}+\lambda\frac{m-1}{2m}\biggr(\frac{\chi}{M^{2}}\biggl)^{m}\biggl),\label{D dim Starobisnky + correction in dual form including chi}
\end{align}
Thus we have to solve Eq.(\ref{chi equation to solve}) to obtain the potentials.
However it is difficult to solve the equation for general values of $n$ and $m$. Thus we use successive iteration to calculate the approximate solution of Eq. (\ref{chi equation to solve}), assuming that $\lambda$ is sufficiently small,
\begin{align}
\frac{\chi}{M^{2}}&=\biggr[(e^{\alpha\phi}-1)-\lambda\biggr(\frac{\chi}{M^{2}}\biggl)^{m-1}\biggl]^{\frac{1}{n-1}}\nonumber\\
&=(e^{\alpha\phi}-1)^{\frac{1}{n-1}}\biggr[1-\frac{\lambda}{n-1}(e^{\alpha\phi}-1)^{\frac{m-n}{n-1}}+\mathcal{O}\Bigr((\lambda(e^{\alpha\phi}-1)^{\frac{m-n}{n-1}})^{2}\Bigl)\biggl].\label{chi approximate solution}
\end{align}
We assume $|\lambda(e^{\alpha\phi}-1)^{\frac{m-n}{n-1}}|\ll 1$ as a condition for convergence of the series. Substituting Eq.(\ref{chi approximate solution}) into Eq.(\ref{D dim Starobisnky + correction in dual form including chi}), we obtain the following potential, 
\begin{align}
V(\phi)&=\frac{n-1}{2n}M_{P}^{2}M^{2}(1-e^{-\alpha\phi})^{\frac{n}{n-1}}\biggr[1-\lambda\frac{n}{m(n-1)}(e^{\alpha\phi}-1)^{\frac{m-n}{n-1}}+\mathcal{O}\Bigr((\lambda(e^{\alpha\phi}-1)^{\frac{m-n}{n-1}})^{2}\Bigl)\biggl]\nonumber\\
&\equiv V_{0}(\phi)\Bigl(1-\lambda \delta V(\phi)+\mathcal{O}((\lambda \delta V)^2)\Bigr),\label{corrected D dim Starobinsky potential}
\end{align}
where $V_{0}$ is $D$-dimensional Starobinsky potential Eq.(\ref{D dim Starobinsky potential}) and $\lambda \delta V$ is a leading correction derived from the additional term $\lambda R^{m}/mM^{2m-2}$. From the shape of the correction term, we can find that  if $m>n$, the potential Eq.(\ref{corrected D dim Starobinsky potential}) becomes close to $V_{0}$ at large field region. On the other hands, we can see that if $m<n$, the correction term becomes large at large field region and perturbation condition for successive iteration will be broken at sufficiently large field region. Also the correction term pushes the potential lower (upper) when $\lambda>0$ ($\lambda<0$). The above statements can be confirmed explicitly by drawing  the potential shape (Fig.\ref{modified potential shape}).  
\par
From here on, we will discuss inflation using approximate potential Eq.(\ref{corrected D dim Starobinsky potential}), assuming $1\ll \phi$ but $|\lambda e^{\frac{m-n}{n-1}\alpha\phi}|\ll 1$ region. Potential slow-roll parameters can be evaluated as follows:
\begin{align}
\epsilon_{V}(\phi)&\simeq\frac{n-1}{2n-1}\biggr[\frac{n}{n-1}e^{-\alpha\phi}-\lambda \frac{n(m-n)}{m(n-1)^{2}}e^{\frac{m-n}{n-1}\alpha\phi}+\mathcal{O}\Bigl((\lambda e^{\frac{m-n}{n-1}\alpha\phi})^2\Bigr)\biggl]^{2},\label{epsilon}\\
\eta_{V}(\phi)&\simeq\frac{2n-2}{2n-1}\biggr[-\frac{n}{n-1}e^{-\alpha\phi}-\lambda\frac{n(m-n)^{2}}{m(n-1)^{3}}e^{\frac{m-n}{n-1}\alpha\phi}+\mathcal{O}\Bigl((\lambda e^{\frac{m-n}{n-1}\alpha\phi})^2\Bigr)\biggl].\label{eta}
\end{align}
Also the e-folding number can be evaluated as follows:
\begin{align}
N_{e}(\phi)\simeq\frac{2n-1}{2n}e^{\alpha\phi}F\biggr(\frac{n-1}{m-1},1,1+\frac{n-1}{m-1},\lambda \frac{m-n}{m(n-1)}e^{\frac{m-1}{n-1}\alpha\phi}\biggl)\times(1+\mathcal{O}(\lambda e^{\frac{m-n}{n-1}\alpha\phi})).\label{e-folding number}
\end{align}
where F(a,b,c,z) denotes Gauss' hyper-geometric function. The Gauss' hyper-geometric function has divergence at $1=\lambda \frac{m-n}{m(n-1)}e^{\frac{m-1}{n-1}\alpha\phi}\Leftrightarrow\phi=\frac{1}{\alpha}\frac{n-1}{m-1}\log\frac{m(n-1)}{\lambda(m-n)}$ if $\lambda(m-n)>0$. It is not surprising because the potential Eq.(\ref{corrected D dim Starobinsky potential}) has maximum value at the field value (see Fig.\ref{modified potential shape}). Such divergence of e-folding number also appears in hilltop type inflation models\cite{Boubekeur and Lyth'05}. 
\par
For further calculation, we assume $|\lambda e^{\frac{m-1}{n-1}\alpha\phi}|\ll 1$. Thus we can expand Gauss' hyper-geometric function and solve Eq.(\ref{e-folding number}) for $\phi$ perturbatively as follow:
\begin{align}
e^{\alpha\phi(N_{e})}=\frac{2n}{2n-1}N_{e}\biggr(1-\lambda\frac{m-n}{m(m+n-2)}\biggr(\frac{2n}{2n-1}N_{e}\biggl)^{\frac{m-1}{n-1}}+\mathcal{O}\Bigr((\lambda N_{e}^{\frac{m-1}{n-1}})^{2}\Bigl)\biggl).\label{horizon crossing phi}
\end{align}
Note that $|\lambda e^{\frac{m-1}{n-1}\alpha\phi}|\ll 1$ is a stronger condition than $|\lambda e^{\frac{m-n}{n-1}\alpha\phi}|\ll 1$ at large field region ($\because \lambda e^{\frac{m-1}{n-1}\alpha\phi}=\lambda e^{\frac{m-n}{n-1}\alpha\phi}\times e^{\alpha\phi}$).
\par
From Eqs.(\ref{epsilon})(\ref{eta})(\ref{horizon crossing phi}), we can obtain the following spectral index $n_{s}$ and  tensor-to-scalar ratio $r$:
\begin{align}
n_{s}&\simeq 1-6\epsilon_{V} (\phi(N_{e}))+2\eta_{V}(\phi(N_{e}))+\cdots\nonumber\\ 
&=1-\frac{2}{N_{e}}\biggr[1+\lambda\frac{(m-1)^2(m-n)}{m(n-1)^{2}(m+n-2)}\biggr(\frac{2n}{2n-1}N_{e}\biggl)^{\frac{m-1}{n-1}}+\mathcal{O}\Bigr((\lambda N_{e}^{\frac{m-1}{n-1}})^{2}\Bigl)\biggl]+\cdots,\label{polynomial spectral index}\\
r&\simeq16\epsilon_{V}(\phi(N_{e}))+\cdots\nonumber\\
&=\frac{4(2n-1)}{n-1}N_{e}^{-2}\biggr[1-2\lambda\frac{(m-1)(m-n)}{m(n-1)(m+n-2)}\biggr(\frac{2n}{2n-1}N_{e}\biggl)^{\frac{m-1}{n-1}}+\mathcal{O}\Bigr((\lambda N_{e}^{\frac{m-1}{n-1}})^{2}\Bigl)\biggl]+\cdots,\label{polynomial tensor-to-scalar ratio}
\end{align}
where $\cdots$ denotes higher order contributions of the slow-roll parameters we neglect here. Therefore there is non-negligible correction $\mathcal{O}(\lambda N_{e}^{\frac{m-1}{n-1}})$  in $n_{s}$ and $r$ derived from $\lambda R^{m}/mM^{2m-2}$ term if it surpasses the higher order contribution of the slow-roll parameters. In this case, leading terms of variations of $n_{s}$ and $r$ from $D$-dimensional Starobinsky model are written as follow:
\begin{align}
\delta n_{s}&=-4\lambda\frac{n(m-1)^{2}(m-n)}{m(n-1)^{2}(2n-1)(m+n-2)}\biggr(\frac{2n}{2n-1}N_{e}\biggl)^{\frac{m-n}{n-1}},\label{variation of spectral index}\\
\delta r&=-32\lambda\frac{n^{2}(m-1)(m-n)}{m(n-1)^{2}(2n-1)(m+n-2)}\biggr(\frac{2n}{2n-1}N_{e}\biggl)^{\frac{m-n}{n-1}-1}.\label{variation of tensor-to-scalar ratio}
\end{align}
These expressions reproduce the results of previous research when $n=2$.
Note that although we considered only a single additional term to $D$-dimensional Starobinsky action, one can expect that if we consider a number of additional terms like $\sum_{m\neq n}\lambda_{m} R^{m}/mM^{2m-2}$, these contributions appear as linear sum of Eq.(\ref{variation of spectral index})(\ref{variation of tensor-to-scalar ratio}) at leading order.
\par
From Eqs.(\ref{variation of spectral index})(\ref{variation of tensor-to-scalar ratio}), we can realize the following facts immediately.
\begin{itemize}
\item The sign of $\lambda(m-n)$ determines the sign of $\delta n_{s}$ and $\delta r$.
\par
$\delta n_{s}<0$ and $\delta r<0$ ($\delta n_{s}>0$ and $\delta r>0$) if $\lambda(m-n)>0$ ($\lambda(m-n)<0$). In the case where we add a number of additional terms to Starobinsky action, we can expect that corrections are partially canceled out by each other at leading order if signs of $\lambda_{m}(m-n)$ are different among the additional terms. Even if $\forall \lambda_{m}>0$, such cancellation occurs as long as $n>2$.
\item $n$ and $m$ determine the power of $N_{e}$ in $\delta n_{s}$ and $\delta r$. 
\par
$\delta n_{s}$ has a positive (negative) power of $N_{e}$ if $m>n$ ($m<n$). And $\delta r$ has a positive (negative) power of $N_{e}$ if $m>2n-1$ ($m<2n-1$). In both cases, the negative power corrections do not have large contributions because $1\ll N_{e}$.
\end{itemize}
For further consideration, we set $m=n+1$ to simplify Eq.(\ref{variation of spectral index})(\ref{variation of tensor-to-scalar ratio}),
\begin{align}
\delta n_{s}&=-4\lambda\frac{n^3}{(n+1)(n-1)^{2}(2n-1)^{2}}\biggr(\frac{2n}{2n-1}N_{e}\biggl)^{\frac{1}{n-1}}\sim-\lambda \mathcal{O}(n^{-2})N_{e}^{\frac{1}{n-1}},\label{ns variation m=n+1}\\
\delta r&=-32\lambda\frac{n^3}{(n+1)(n-1)^{2}(2n-1)^{2}}\biggr(\frac{2n}{2n-1}N_{e}\biggl)^{\frac{1}{n-1}-1}\sim-\lambda \mathcal{O}(n^{-2})N_{e}^{\frac{1}{n-1}}\times\frac{8}{N_{e}}.\label{r variation m=n+1}
\end{align}
From Eqs.(\ref{ns variation m=n+1})(\ref{r variation m=n+1}), we can realize that the leading term of variations $\delta n_{s}$ and $\delta r$ becomes smaller as we consider larger $n$ (i.e. larger $D$ because $D=2n$). Considering that $\lambda$ is restricted by Planck2018  observational constraints of $n_{s}$ and $r$, this imply that the restriction of $\lambda$ can be relaxed in higher-dimensional Starobinsky model.
\subsection{Numerical approach}
In this subsection, we will calculate $n_{s}$ and $r$ numerically and obtain a constraint of $\lambda$ form CMB observation. The scheme of the numerical calculation is summarized as follows.
\begin{itemize}
\item[(i)] Find a solution $\chi^{(k)}$ of Eq.(\ref{chi equation to solve}) by $k$ times successive iteration. Here we assume $|\lambda(e^{\alpha\phi}-1)^{\frac{m-n}{n-1}}|< 1$ for convergence of a series of the solution. 
\item[(ii)] Calculate spectral index $n_{s}^{(k)}$ and tensor-to-scalar ratio $r^{(k)}$ by using the solution $\chi^{(k)}$. Here we do NOT make any assumption like $1\ll \phi$ or $|\lambda e^{\frac{m-1}{n-1}\alpha\phi}|\ll 1$.
\item[(iii)]Output $n_{s}^{(k)}$ and $r^{(k)}$ as results, if $n_{s}^{(k)}$ and $r^{(k)}$ satisfy the following conditions, otherwise try again from (i) by increasing $k$:
\begin{align}
n_{s}^{(k-1)}-n_{s}^{(k)}<\frac{1}{N_{e}^2}\hspace{5em}r^{(k-1)}-r^{(k)}<\frac{1}{N_{e}^3}.\label{condition of numerical calculation}
\end{align}
\item[(iv)]Repeat (i)(ii)(iii) varying e-folding number $N_{e}$ and parameter $\lambda$.
\end{itemize}
\par
\begin{figure}
\centering
\includegraphics[width=10cm]{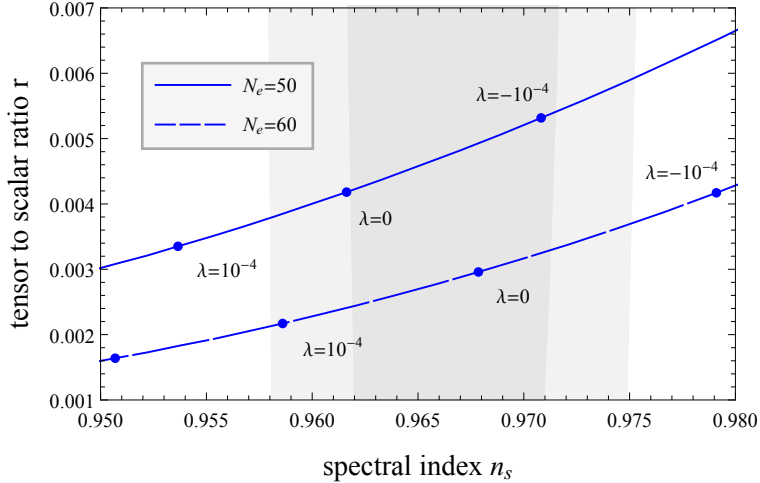}
\caption{This figure shows numerical predictions of $n_{s}$-$r$ plot in 4-dimensional Starobinsky model with $\lambda R^3/3M^4$ term as we vary e-folding number $N_{e}$ and parameter $\lambda$. The gray region denotes constraints from PlanckTT+lowP+BKP+lensing+ext experiments, which is traced from Fig.18 in Planck2018 results}
\label{ns-r plot of R+R^2+R^3 in 4D}
\end{figure}
\begin{figure}
\centering
\includegraphics[width=10cm]{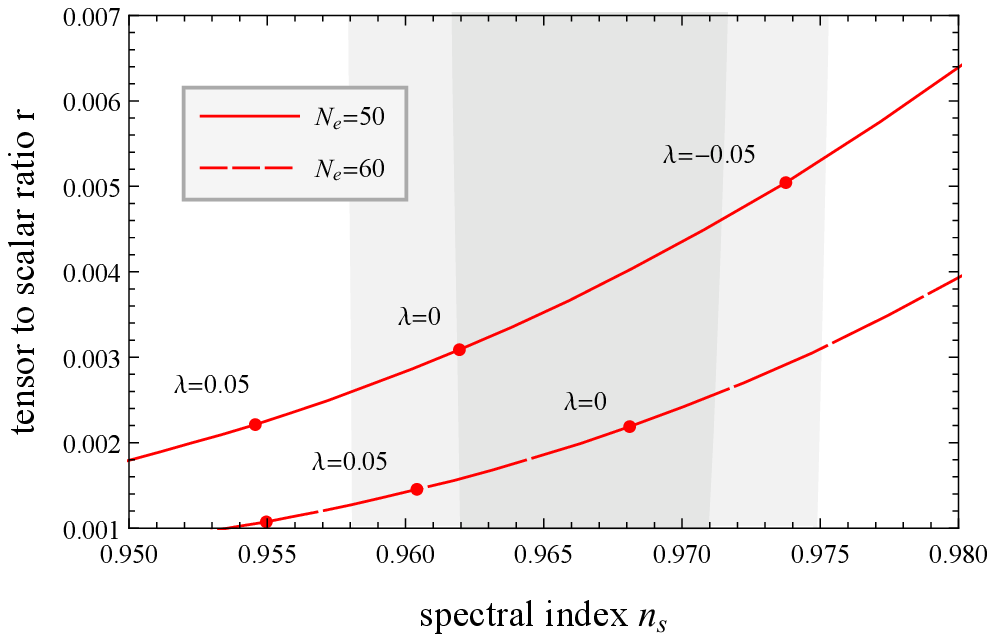}
\caption{This figure shows numerical predictions of $n_{s}$-$r$ plot in 10-dimensional Starobinsky model with $\lambda R^6/6M^{10}$ term as we vary e-folding number $N_{e}$ and parameter $\lambda$. The gray region denotes constraints from PlanckTT+lowP+BKP+lensing+ext experiments, which is traced from Fig.18 in Planck2018 results.}
\label{ns-r plot of R+R^5+R^6 in 10D}
\end{figure}
\begin{figure}
\centering
\includegraphics[width=10cm]{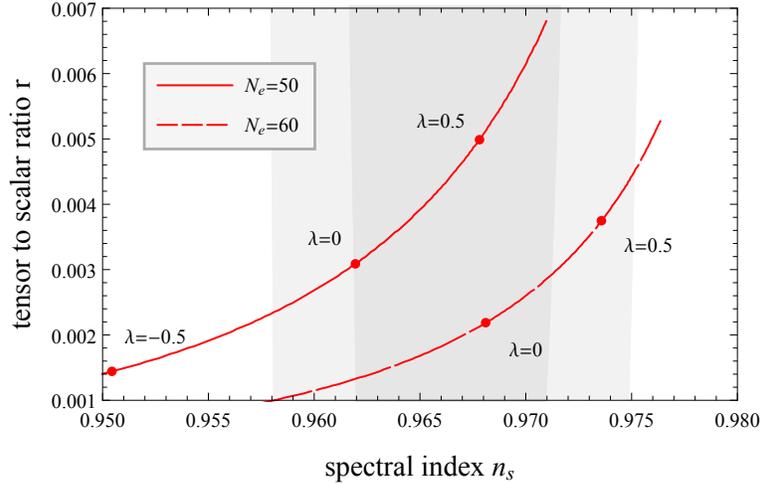}
\caption{This figure shows numerical predictions of $n_{s}$-$r$ plot in 10-dimensional Starobinsky model with $\lambda R^4/4M^{6}$ term as we vary e-folding number $N_{e}$ and parameter $\lambda$. The gray region denotes constraints from PlanckTT+lowP+BKP+lensing+ext experiments, which is traced from Fig.18 in Planck2018 results.}
\label{ns-r plot of R+R^5+R^4 in 10D}
\end{figure}

In this calculation, we assume only $|\lambda(e^{\alpha\phi}-1)^{\frac{m-n}{n-1}}|< 1$, whereas we assumed $|\lambda(e^{\alpha\phi}-1)^{\frac{m-n}{n-1}}|\ll 1$,  $1\ll \phi$ and $|\lambda e^{\frac{m-1}{n-1}\alpha\phi}|\ll 1$ in previous analytical calculations. Thus this numerical calculation is valid in a broader range than analytical one. Eq.(\ref{condition of numerical calculation}) are conditions for an error from successive iteration not to surpass an error from higher order of slow-roll parameters. Further accuracy is not needed because we neglect higher order of slow-roll parameters here. This condition can be satisfied for large $k$ as long as $|\lambda(e^{\alpha\phi}-1)^{\frac{m-n}{n-1}}|< 1$. Fig.\ref{ns-r plot of R+R^2+R^3 in 4D}, \ref{ns-r plot of R+R^5+R^6 in 10D} and \ref{ns-r plot of R+R^5+R^4 in 10D} show numerical predictions calculated through the steps above. Fig.\ref{ns-r plot of R+R^2+R^3 in 4D} shows numerical predictions of $n_{s}$-$r$ plot in 4-dimensional Starobinsky model with $\lambda R^3/3M^4$ terms. And Fig.\ref{ns-r plot of R+R^5+R^6 in 10D} and Fig.\ref{ns-r plot of R+R^5+R^4 in 10D} shows numerical predictions of $n_{s}$-$r$ plot in 10-dimensional Starobinsky model with $\lambda R^6/6M^{10}$ and  $\lambda R^4/4M^6$ term, respectively. 
\par 
\par
In Fig.\ref{ns-r plot of R+R^5+R^4 in 10D}, the lines of the prediction are broken off because the condition (\ref{condition of numerical calculation}) can not be satisfied even for large $k$ (we performed the numerical calculation until $k=100$). This implies that $|\lambda(e^{\alpha\phi}-1)^{\frac{m-n}{n-1}}|< 1$ is broken, i.e., the perturbation is broken. Therefore we stop the calculation in further range because it is impossible to apply this method in $|\lambda(e^{\alpha\phi}-1)^{\frac{m-n}{n-1}}|> 1$. We would like to comment, however, that it is possible to calculate in a further range if one can solve Eq.(\ref{chi equation to solve}) rigidly. 
\par
For further investigation, we calculate a constraint of $\lambda$ in 4 and 10-dimensional Starobinsky model with $R^{m}/mM^{2m-2}$ term ($2\le m\le10$ and $m\neq n$) from Planck2018 results, $n_{s}=0.9665\pm0.0038$ (68\%CL PlanckTT,TE,EE+lensing+BAO). Table.\ref{table of lambda} shows results of the calculations. $(PB)$ denotes that we could not obtain upper or lower bounds of $\lambda$ because the perturbation is broken (we performed the calculation until $k=100$ as before). As mentioned before, it is possible to calculate the bounds if one can solve Eq.(\ref{chi equation to solve}) rigidly. 
\begin{table}[htb]
\centering
\begin{tabular}{|c||c|c|}\hline
					&$D=2n=4$																	&$D=2n=10$																			\\\hline
\multirow{2}{*}{$m=2$}	&\multirow{2}{*}{(4D Starobinsky term)}											&$\hspace{3em}(PB)\lesssim\lambda\lesssim(PB)\hspace{4.5em}(N_{e}=50)$	\\
					&																			&$\hspace{3.25em}-0.9\lesssim\lambda\lesssim(PB)\hspace{4.5em}(N_{e}=60)$			\\\hline
\multirow{2}{*}{$m=3$}	&$\hspace{0.5em}-9.5\times10^{-5}\lesssim\lambda\lesssim-1.3\times10^{-5}\hspace{1.25em}(N_{e}=50)$	&$+1.1\times10^{-1}\lesssim\lambda\lesssim(PB)\hspace{4.5em}(N_{e}=50)$				\\
					&$\hspace{0.5em}-2.3\times10^{-5}\lesssim\lambda\lesssim+5.4\times10^{-5}\hspace{1.25em}(N_{e}=60)$	&$-4.7\times10^{-1}\lesssim\lambda\lesssim+4.9\times10^{-1}\hspace{1.5em}(N_{e}=60)$				\\\hline
\multirow{2}{*}{$m=4$}	&$\hspace{0.5em}-5.3\times10^{-7}\lesssim\lambda\lesssim-0.7\times10^{-7}\hspace{1.25em}(N_{e}=50)$	&$+0.5\times10^{-1}\lesssim\lambda\lesssim +8.4\times 10^{-1}\hspace{1.5em}(N_{e}=50)$				\\
					&$\hspace{0.5em}-1.1\times10^{-7}\lesssim\lambda\lesssim+2.5\times10^{-7}\hspace{1.25em}(N_{e}=60)$	&$-2.9\times10^{-1}\lesssim\lambda\lesssim +1.6\times 10^{-1}\hspace{1.5em}(N_{e}=60)$				\\\hline
\multirow{2}{*}{$m=5$}	&$\hspace{0.5em}-4.3\times10^{-9}\lesssim\lambda\lesssim-0.6\times10^{-9}\hspace{1.25em}(N_{e}=50)$	&\multirow{2}{*}{(10D Starobinsky term)}														\\
					&$\hspace{0.5em}-0.8\times10^{-9}\lesssim\lambda\lesssim+1.7\times10^{-9}\hspace{1.25em}(N_{e}=60)$	&																						\\\hline
\multirow{2}{*}{$m=6$}	&$-4.2\times10^{-11}\lesssim\lambda\lesssim-0.6\times10^{-11}\hspace{1em}(N_{e}=50)$	&$-3.8\times10^{-2}\lesssim\lambda\lesssim-0.4\times10^{-2}\hspace{1.5em}(N_{e}=50)$				\\
					&$-0.6\times10^{-11}\lesssim\lambda\lesssim+1.4\times10^{-11}\hspace{1em}(N_{e}=60)$	&$-1.1\times10^{-2}\lesssim\lambda\lesssim+3.3\times10^{-2}\hspace{1.5em}(N_{e}=60)$				\\\hline
\multirow{2}{*}{$m=7$}	&$-4.5\times10^{-13}\lesssim\lambda\lesssim-0.6\times10^{-13}\hspace{1em}(N_{e}=50)$	&$-6.4\times10^{-3}\lesssim\lambda\lesssim-0.7\times10^{-3}\hspace{1.5em}(N_{e}=50)$				\\
					&$-0.5\times10^{-13}\lesssim\lambda\lesssim+1.2\times10^{-13}\hspace{1em}(N_{e}=60)$	&$-1.7\times10^{-3}\lesssim\lambda\lesssim+5.0\times10^{-3}\hspace{1.5em}(N_{e}=60)$				\\\hline
\multirow{2}{*}{$m=8$}	&$-5.1\times10^{-15}\lesssim\lambda\lesssim-0.6\times10^{-15}\hspace{1em}(N_{e}=50)$	&$-1.4\times10^{-3}\lesssim\lambda\lesssim-0.2\times10^{-3}\hspace{1.5em}(N_{e}=50)$				\\
					&$-0.5\times10^{-15}\lesssim\lambda\lesssim+1.1\times10^{-15}\hspace{1em}(N_{e}=60)$	&$-0.4\times10^{-3}\lesssim\lambda\lesssim+0.1\times10^{-3}\hspace{1.5em}(N_{e}=60)$				\\\hline
\end{tabular}
\caption{This table shows $\lambda$ constraint in 4- and 10-dimensional Starobinsky model with $R^{m}/mM^{2m-2}$ term ($2\le m\le10$ and $m\neq n$) from Planck results. Here we set $N_{e}=50,60$.}
\label{table of lambda}
\end{table}
\par
As all figures and table show, we can conclude that $\lambda$ constraint in higher-dimensional Starobinsky model with $\lambda R^{m}/mM^{2m-2}$ term ($n\neq m$) is significantly relaxed in higher dimension. This numerical results is also supported by the implication of analytical results in the previous subsection. Therefore, although we calculate only $D=4,10$ cases in this paper, one can expect that this relaxation globally happens in higher dimensions.  
\section{Summary and discussion}\label{conclu}
4-dimensional Starobinsky model, whose action has a curvature squared terms $R^2/2M^2$, is one of the most promising inflation models. However, the origin of the higher curvature term is still unknown. From the viewpoint of higher curvature extensions of Einstein gravity, there is no reason to exclude $R^{m} (2<m)$ terms. In Ref.\cite{Qing-Guo Huang'14} the authors added a $\lambda R^{m}/mM^{2m-2} (2<m)$ term to the 4-dimensional Starobinsky action and estimated its effects on observational predictions. They obtained a conclusion that some observables are quite sensitive to the existence of $\lambda R^{m}/mM^{2m-2} (2<m)$ terms.
\par
In this paper, we extended the analysis of Ref.\cite{Qing-Guo Huang'14} to $D$-dimensional Starobinsky model, motivated by recent works on higher-dimensional Starobinsky model \cite{Ketov and Nakata'17a}\cite{Otero et al'17}\cite{Ketov and Nakata'17b}. This extension is reasonable because an effective action of high energy physics, which contains higher curvature terms, is expected to appear as Eq.(\ref{a series of curvature}). 
\par
First, we considered models which have $R+R^n/nM^{2n-2}$ action in $D$ dimension and clarified that when and only when $D=2n$ is satisfied, the model can cause a successful inflation. Then we added a $\lambda R^{m}/mM^{2m-2}$ $(m\neq n)$ term to $D=2n$-dimensional Starobinsky action and estimated its effects on observational predictions in both analytical and numerical ways. In the analytical approach, we find that the deviations of observables caused by the additional terms become smaller in higher-dimensions. Also we have checked the result from the numerical approach. Therefore we can conclude that the observational predictions of $D$-dimensional ($4<D$) extended Starobinsky model are less sensitive to such terms than those of the original 4-dimensional model. This result make it easier to construct Starobinsky-like models in higher dimensions, that is a desired future from a viewpoint of the unified description of fundamental forces based on, e.g., supergravity/strings.
\par
As a final remark, we have to come back to the discussion below Eq.(\ref{Starobinsky action}). From the observations, there must be a hierarchy between $M$ and $M_{P}$, i.e., $M\sim \mathcal{O}(10^{-5})\times M_{P}$. One might think that $M\sim M_{(D)}$ can be set by tuning $\mathcal{V}_{\rm{extra}}$ because $M_{P}^{2}=M_{(D)}^{D-2}\mathcal{V}_{\rm{extra}}$. However, we find the following constraint assuming a condition that KK massive modes are decoupled during inflation $M\lesssim1/\mathcal{V}_{\rm{extra}}^{1/D-4}$;
\begin{align}
M \lesssim\mathcal{O}(10^{-\frac{5D}{2D-4}})\times M_{(D)}.\label{mass hierarchy in D}
\end{align}
In higher dimension ($4<D$), this hierarchy is milder than 4-dimensional one. Nevertheless there still exists a milder hierarchy. To avoid this remained hierarchy, we may have to introduce another scale, such as a brane tension, or consider the non-trivial compactification mechanism.
\begin{acknowledgments}
Y.A. would like to thank Hiroyuki Abe and Shuntaro Aoki for useful discussion and comments. 
\end{acknowledgments}
\appendix
\section{F(R) gravity  and Legendre-Weyl transformation in arbitrary dimension}\label{App}
Let us assume the following $F(R)$ gravity action in $D$ dimension:
\begin{align}
\frac{M_{(D)}^{D-2}}{2}\int d^Dx\sqrt{-g}\Bigl(R+F(R)\Bigr),\label{appendix F(R) gravity}
\end{align}
where F(R) is an arbitrary function of Ricci scalar at this point. We recast Eq.(\ref{appendix F(R) gravity}) using auxiliary field $\chi$ as follow:
\begin{align}
\frac{M_{(D)}^{D-2}}{2}\int d^Dx\sqrt{-g}\Bigl(R+F(\chi)+F'(\chi)(R-\chi)\Bigr),\label{appendix rewrite by auxiliary field}
\end{align}
where $'$ denotes derivative with respect to $\chi$. If one varies Eq.(\ref{appendix rewrite by auxiliary field}) with respect to $\chi$, one can realize that equation of motion has simply form $\chi=R$ (here $F''(\chi)\neq0$ is assumed). Substitution of this algebraic equation turns Eq.(\ref{appendix rewrite by auxiliary field}) into original Eq.(\ref{appendix F(R) gravity}). In this meaning, one can say Eq.(\ref{appendix F(R) gravity}) and Eq(\ref{appendix rewrite by auxiliary field}) are equivalent.
\par
Eq.(\ref{appendix rewrite by auxiliary field}) is non-minimally coupled scalar tensor action without kinetic term of scalar field. It is well known that such action can be rewritten into Einstein-Hilbert action with minimally coupled scalar field by the following scalar field dependent metric redefinition:
\begin{align}
g^{AB}=\Bigl(F'(\chi)+1\Bigr)^{\frac{2}{D-2}}g^{AB}_{(E)},\label{appendix redefinition of metric}
\end{align}
where $g^{AB}_{(E)}$ is redefined metric. Using Eq(\ref{appendix redefinition of metric}) one can calculate as follow: 
\begin{align}
\sqrt{-g}&=\Bigl(F'(\chi)+1\Bigr)^{-\frac{D}{D-2}}\sqrt{-g_{(E)}},\label{appendix determinant transformation}\\
R&=\Bigl(F'(\chi)+1\Bigr)^{\frac{2}{D-2}}\Bigl[R_{(E)}-\frac{D-1}{D-2}g_{(E)}^{AB}\partial_{A}\ln (F'(\chi)+1)\partial_{B}\ln (F'(\chi)+1)+\cdots\Bigr],\label{appendix Ricci scalar transformation}
\end{align}
where $g_{(E)}$ and $R_{(E)}$ is determinant and Ricci scalar which are constructed from $g^{AB}_{(E)}$. Also $\cdots$ denotes total derivative terms, which we will neglect later. Substituting Eq.(\ref{appendix determinant transformation}) and Eq.(\ref{appendix Ricci scalar transformation}) into Eq.(\ref{appendix rewrite by auxiliary field}), we obtain the following action:
\begin{align}
&\frac{M_{(D)}^{D-2}}{2}\int d^Dx\sqrt{-g_{(E)}}\Bigl[R_{(E)}-\frac{D-1}{D-2}g^{AB}_{(E)}\partial_{A}\bar{\phi}\partial_{B}\bar{\phi}\nonumber\\
&\hspace{15em}-\Bigl(F'(\chi)+1\Bigr)^{-\frac{D}{D-2}}\Bigl(F'(\chi)\chi-F(\chi)\Bigr)\Bigr],
\end{align}
where $\bar{\phi}\equiv\ln\Bigl(F'(\chi)+1\Bigr)$. We have to remark that one has to solve the equation $F'(\chi)=e^{\bar{\phi}}-1$ with respect to $\chi$ in order to obtain a canonical kinetic term.
If one chooses a function form of $F(R)$ as $R^{n}/nM^{2n-2}$ or $R^{n}/nM^{2n-2}+\lambda R^{m}/mM^{2m-2}$, one obtains Eq.(\ref{D dim Starobinsky in D dim dual form}) or Eq.(\ref{dual form of D-dimensional Starobinsky model with correction}), respectively.

\end{document}